\documentclass[12pt,preprint]{aastex}

\def\Msol{\thinspace\hbox{$\hbox{M}_{\odot}$}}

\def\a4{\hsize 17.0cm \vsize 25.cm}

\shorttitle{X-ray Emission from SSCs and their Superbubbles}

\shortauthors{Silich et al.}

\begin{document}

\title{On the X-ray Emission from Massive Star Clusters and 
       their Evolving Superbubbles }

\author{
Sergiy Silich, Guillermo  Tenorio-Tagle, Gabriel Alejandro A\~norve Zeferino}
\affil{Instituto Nacional de Astrof\'\i sica Optica y
Electr\'onica, AP 51, 72000 Puebla, M\'exico; silich@inaoep.mx}

\begin{abstract}
Here we discuss the X-ray emission properties from the hot
thermalized plasma that results from the collisions of individual stellar 
winds and supernovae ejecta within rich
and compact star clusters. We propose a simple analytical way of
estimating the X-ray emission generated by super star clusters and
derive an expression that indicates how this X-ray emission depends
on the main cluster parameters. Our model 
predicts that the X-ray luminosity from the star cluster region is 
highly dependent on the star cluster wind terminal speed, a quantity
related to the temperature of the thermalized ejecta.
We have also compared the X-ray luminosity from the SSC plasma with the 
luminosity of the interstellar bubbles generated from the mechanical
interaction of the high velocity star cluster winds with the ISM.
We found that the hard (2.0 keV - 8.0 keV) X-ray emission is usually 
dominated by the hotter SSC plasma whereas the soft (0.3 keV - 2.0
keV) component is dominated by the bubble plasma.
This implies that compact and massive star clusters should
be detected as point-like hard X-ray sources embedded into extended 
regions of soft diffuse X-ray emission. We also compared our results
with predictions from the population synthesis models that take
into consideration binary systems and found that in the case of young,
massive and compact super star clusters the X-ray emission from the
thermalized star cluster plasma may be comparable or even larger than
that expected from the HMXB population.
\end{abstract}

\keywords{Star clusters: winds, bubbles, X-ray emission}

\section{Introduction}

In many starburst and in interacting and merging galaxies a substantial 
fraction of the star formation is concentrated in a number of compact, 
young and massive stellar clusters or super stellar clusters (SSCs; 
see, for example, Holtzman et al. 1992; Ho, 1997; O'Connell, 2004 
and references therein).
The high stellar densities, the large  energy and mass 
deposition rates,  provided by stellar winds and supernovae explosions (SNe), 
suggest that SSCs are potentially strong X-ray emitters (Chevalier, 1992). 

Indeed the X-ray emission from NGC 3603, Arches cluster in our Galaxy and
R136 in the LMC, the local analogies to low mass SSCs, has been detected 
(Moffat et al, 2002; Yusef-Zadeh et al. 2002; Stevens \& Hartwell
2003). The X-ray emission
from Arches cluster has a diffuse component and five discrete 
sources. Overall, 40\% of the observed emission is diffuse and 
60\% comes from the discrete sources. The absorption-corrected 
X-ray luminosity (between 0.5 - 8 
keV) is $L_X \approx (0.5 - 3) \times 10^{35}$ erg
s$^{-1}$ and the best spectral fit requires the metallicity of
the X-ray plasma to be 4 - 5 times solar (Law \& Yusef-Zadeh, 2004).
Similarly, the X-ray luminosity from R136 in 30 Doradus is $L_X = 5.5 \times 
10^{34}$ erg s$^{-1}$ (Stevens \& Hartwell 2003). 
On the other hand, an examination of the XMM-Newton observations led 
Smith \& Wang (2004) to the discovery of the
diffuse X-ray emission from a 100 pc superbubble, 30 Doradus C, 
associated with a rich OB-association LH 90, located south-west of 
the  center of 30 Doradus. The examination of the X-ray spectrum 
indicates also an overabundance of $\alpha$-elements in the bubble plasma  
by a factor $\sim 3$, compared to the metallicity
of the LMC.

X-ray emission detected in distant galaxies presents overall  two
components: an extended diffuse component that seems to be related to
a collection of superbubbles associated to individual star clusters,
and a number of unresolved point-like sources (Summers et al. 2004;
Metz et al. 2004; Smith et al. 2005). In some cases, the X-ray
emission shows good coincidence with the young stellar
clusters (Metz et al. 2004). However, there are also many examples 
of X-ray regions that do not present any 
evidence of active star formation. In other cases the X-ray sources 
seem to be shifted relative to the nearby star clusters and some massive 
clusters seem to emit no X-rays at all (Kaaret et al. 2004).
Thus the origin of the X-ray emission in distant galaxies 
and the dependence of the detected X-ray emission on the parameters of
the embedded star clusters remain controversial. 

Here we discuss how the mechanical energy deposited by stellar
clusters restructures the host galaxy ISM while producing a hot plasma
which should be detected in the X-ray regime (see section 2). In 
sections 3 and 4 we demonstrate that the
mechanical energy from SSCs leads to a two-component plasma model: a high
temperature, overabundant in $\alpha-$elements plasma, ejected from the star 
cluster, and a lower temperature, extended, interstellar bubble component.  
Here we discuss how the luminosity of the hot component depends on the
star cluster parameters and in section 5 we compare the contributions
from both components when evolving in different 
interstellar environments and as a function of time.

\section{Super Star Cluster Winds and their Impact on the ISM}

The interaction of the high velocity outflow that results from the 
energy injected by multiple stellar winds and supernovae (SNe) 
explosions within a star cluster volume, leads to a major
re-structuring of the ISM. The evolution under the assumption of a uniform 
density case, causes, as in the case of a stellar wind (Weaver et al. 
1977), several physically distinct  regions (see Figure 1), promoted
by the shock waves (inner and outer shocks) that result from the 
wind-ISM interaction. The X-ray emission from such structure, must 
take first into account the central star cluster (zone A), filled with the 
thermalized, high temperature plasma, injected by winds and SNe. There 
the large central overpressure that results from thermalization,  
accelerates the injected gas and eventually blows it out of the star 
cluster volume while generating a stationary wind (Chevalier \& Clegg, 
1985). The central zone A, a powerful X-ray emitter, 
is surrounded by the  free-wind region (zone B in Figure 1) where the 
high metallicity matter emanating from the SSC volume rapidly acquires 
its terminal velocity ($v_{A,\infty}$), while its density and temperature 
begin to  steadily drop as $r^{-2}$ and $r^{-4/3}$. Zone B is also,
in the case of SSCs, a source of X-ray emission. Its main
contribution, as shown below, arises from 
regions in the immediate neighborhood of the SSC, although formally 
X-rays are emitted throughout the whole zone down to its outer 
boundary: the reverse inner shock. An exception to this latter issue is 
for massive and compact SSCs whose winds become strongly radiative,
bringing the temperature well below the lower limit for X-ray emission 
($T_{cut} = 5 \times 10^5$ K), long before reaching the inner shock
(Silich et al. 2003, 2004). Whether the SSC wind undergoes cooling or not, it 
eventually crosses the inner shock to be once again thermalized. This 
generates and maintains the hot superbubble (zone C), the reservoir of 
thermal energy that drives the outer shock into the ISM. The hot 
superbubble spans from the location of the inner shock to the contact
discontinuity that separates the hot wind from the expanding outer shell of 
interstellar swept up matter, accelerated by the leading shock 
(zone D). Given the
strength of the shocks, at first the two outer zones are strong X-ray 
emitters. Soon however, particularly for large ISM densitites 
($n_{ISM} \ge 1$ cm$^{-3}$), radiative cooling brings the temperature down 
in zone D, causing its condensation into a thin, dense and cool 
expanding outer shell. In cases in which magnetic fields are not considered
the internal structure of zone C, its chemical composition and its 
X-ray emissivity, depend strongly on the evolution time, $t$, and on 
the rate at which  mass is evaporated from the cold outer shell into
the hot superbubble interior (see Bisnovatyi-Kogan \& Silich, 1995 and
references therein). 
\begin{figure}[htbp]
\plotone{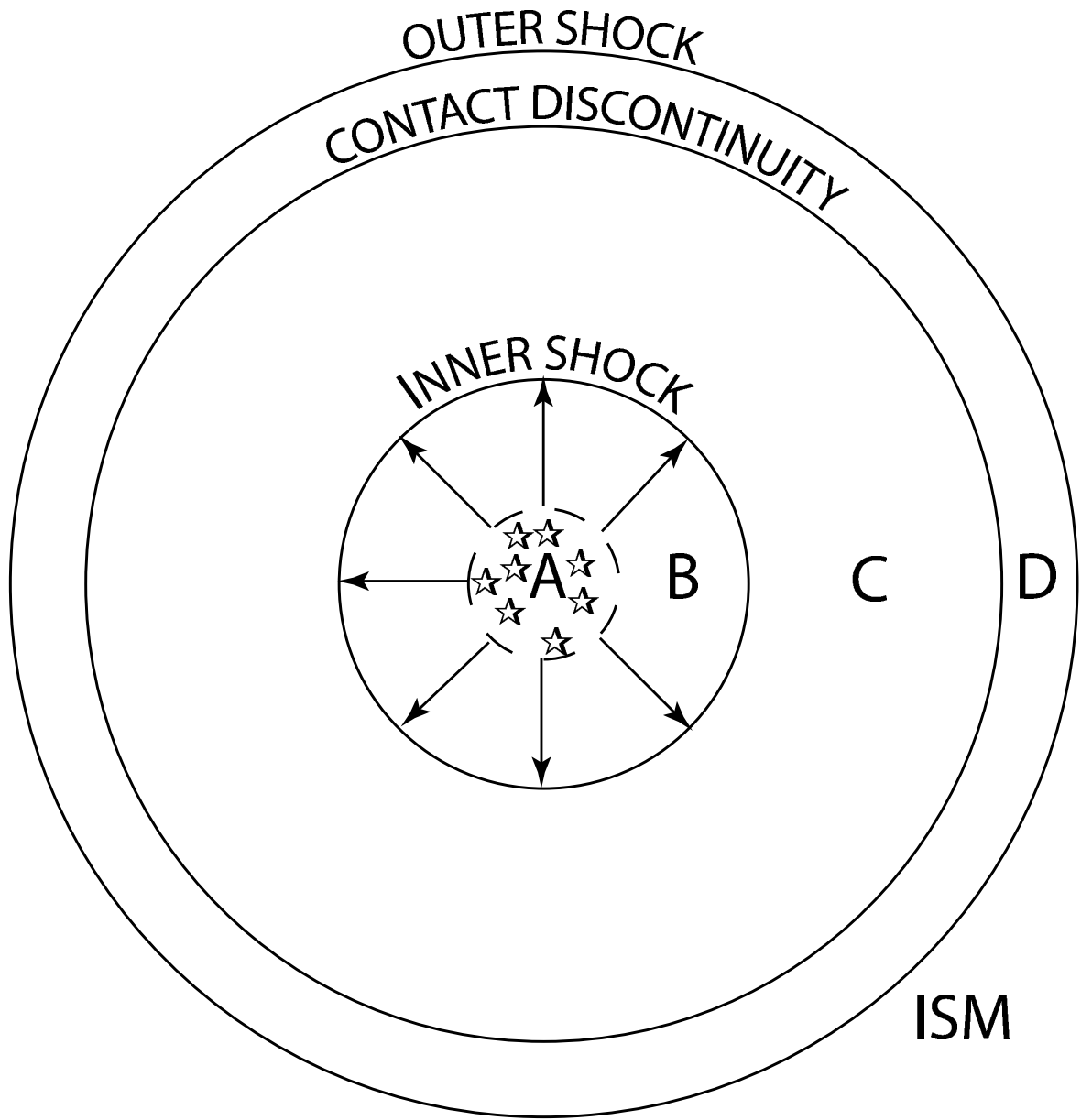}
\caption{SSC winds and the structure of the ISM. Schematic
representation of the structure generated by the mass and energy 
deposition rates within SSCs. The central zone (A) represents the SSC 
volume itself, where stellar winds and SNe from the evolving cluster 
are thoroughly thermalized, causing the large thermal pressure that 
drives the SSC wind. The remaining concentric zones are, as in Weaver 
et al. 1977, (B) the free-wind region, (C) the hot superbubble
and (D) the shell of swept up matter, all surrounded by the unperturbed
ISM. Note that zones A and B produce an insignificant X-ray emission 
in the case of winds produced by single stars (see Weaver et al 1977).}
\label{figo}
\end{figure}

\section{The Inner Structure of the X-ray region}

In the adiabatic solution the temperature and density of the gas 
at the star cluster surface, $r = R_{SC}$, may be inferred from the mass and 
energy conservation laws (see equations 4 and 6 in Silich et al. 2004). 
They are: 
\begin{eqnarray}
      \label{eq.1a}
      & & 
T_{SC} = \frac{2(\gamma-1) \mu_i}{\gamma(\gamma+1) k}
              \frac{L_{SC}}{{\dot M}_{SC}} = \frac{\gamma - 1}
            {\gamma (\gamma+1)} \frac{\mu_i}{k} V^2_{A,\infty} =
         1.1 \times 10^7 V^2_{1000} K ,
      \\[0.2cm] & & 
      \label{eq.1b}
\rho_{SC} = \frac{{\dot M}_{SC}}{4 \pi R_{SC}^2 c_{SC}} =
\left(\frac{\gamma + 1}{\gamma - 1}\right)^{1/2} \frac{L_{SC}}
           {2 \pi R^2_{SC} V^3_{A,\infty}} =
3.3  \times 10^{-24} \frac{L_{38}}{R^2_{SC,1} V^3_{1000}} g \, cm^{-3} ,
\end{eqnarray}
where $L_{SC}$ and ${\dot M}_{SC}$ are the mechanical energy and mass
deposition rates, $k$ is the Boltzmann constant, $\gamma = 5/3$ is the 
ratio of specific heats, $\mu_i = 14/23 m_H$ 
is the mean mass per particle for a fully ionized plasma,
$c_{SC} = (\gamma k T / \mu_i)^{1/2} = [(\gamma - 1)/(\gamma+1)]^{1/2}
\, V_{A,\infty}$, is the sound velocity at the star cluster edge, and 
$V_{A,\infty} = (2 L_{SC}/{\dot M}_{SC})^{1/2}$, is the resultant
wind adiabatic terminal speed, $L_{38}$ is the mechanical luminosity of 
the SSC measured in units of $10^{38}$ erg s$^{-1}$, $R_{SC,1}$ is the 
star cluster radius measured in units of 1 pc, and $V_{1000}$ is the 
wind adiabatic terminal speed in units of 1000 km s$^{-1}$. The temperature 
and density of the gas inside the star cluster are always close to the 
adiabatic values if the stationary wind regime is possible.

On the other hand, from the work of  Chevalier \& Clegg (1985), 
Cant{\'o} et al. (2000), who considered the adiabatic solution for SSC winds, 
and from the recent semi-analytical radiative solution of Silich et al. 
(2004) we know that density and temperature inside of the star cluster 
volume remain almost constant (in the adiabatic case the ratio of the 
temperature at the star cluster surface to the central temperature is
$T_{SC}/T_C = 2/(\gamma+1) = 0.75$) and are completely defined by the 
star cluster parameters ($L_{SC}$, ${\dot M}_{SC}$ and $R_{SC}$). 
This is also the case in the radiative solution, although one has to
take also  the chemical composition of the ejected material into 
account. Thus the density and the temperature distributions in zone 
A may be approximated as
\begin{equation}
      \label{eq.1c}
T(r) \approx T_{SC}, \quad
\rho(r) \approx \alpha_{\rho} \rho_{SC} ,
\end{equation}
where the fiducial coefficient $\alpha_{\rho}$ takes into 
consideration the deviation of density in zone A from the 
homogeneous distribution (the small deviation of temperature from the
surface value leads to negligible changes of the X-ray emissivity).

In zone B all variables rapidly approach their asymptotic values,
$\rho(r) \sim r^{-2}, \, T \sim r^{-4/3}$. In the strongly radiative 
regime the temperature may drop faster and reach values $\sim$ $10^4$K 
at smaller radii. However the contribution of this zone to the total 
X-ray luminosity is small and it arises mainly from a region close to 
the SSC surface (see section 4). 

The expansion of the outer shock is supported by the thermal pressure 
of the thermalized wind (zone C in Figure 1). In the case of a
homogeneous ISM and a 
constant rate of mechanical energy deposited by the central cluster, 
the time evolution of the outer shock radius, $R_{out}$, the reverse
shock radius, $R_{in}$, and of the expansion velocity, $V_{out}$, of 
the outer shell are given by (Koo \& McKee, 1992) 
\begin{equation}
      \label{eq.2a}
R_{out} = 0.88 \left(\frac{L_{SC}}{\rho_{ISM}}\right)^{1/5} t^{3/5} , \quad 
R_{in} = 0.92 \left(\frac{L_{SC}}{\rho_{ISM}}\right)^{3/10}
         \frac{t^{2/5}}{V^{1/2}_{A,\infty}} , \quad 
V_{out}  =  \frac{3}{5} \frac{R_{out}}{t} ,
\end{equation}
where $t$ is the bubble age and $\rho_{ISM}$ is the density of the ISM.  

The shocked ISM in zone D remains adiabatic until the evolution time
becomes larger than its characteristic cooling time scale,
$t_{cool}$, (Mac Low \& McCray, 1988), when
\begin{equation}
      \label{eq.3} 
t \ge t_{cool} = 2.3 Z^{-0.42}_{ISM} \times 10^4  n_{ISM}^{-0.71} 
L_{38}^{0.29} yr ,
\end{equation}
where $Z_{ISM}$ is the ISM gas metallicity measured in the Solar units
and $n_{ISM}$ is the ISM gas number density.
When  $t$ exceeds $t_{cool}$, the gas in zone D becomes thermally 
unstable and collapses into a thin cold expanding outer
shell. After this stage the density and the temperature distributions 
inside the bubble (zone C) are dominated by thermal evaporation of 
matter from the cold shell into the superbubble. If thermal
evaporation is not blocked by magnetic fields, the density and 
the temperature distributions in zone C are (Weaver et al. 1977;
Mac Low \& McCray, 1988)
\begin{equation}
      \label{eq.5a}
n(x) = n_0 (1 - x)^{-2/5}; \quad  T(x) = T_0 (1 - x)^{2/5} ,
\end{equation}
where $x = r / R_{CD}$ is the fractional radius, $R_{CD}$ is the
radius of the contact discontinuity, $n_0$ and $T_0$ are 
functions of the evolutionary time: 
$n_0 = 1.0 \times 10^2 L^{6/35}_{38} n^{19/35}_{ISM} t^{-22/35}$ cm$^{-3}$,
$T_0 = 5.5 \times 10^7 L^{8/35}_{38} n^{2/35}_{ISM} t^{-6/35}$ K, and
$t$ here is measured in year units.

\section{The X-ray model}

Once the distributions of density and temperature in the various zones 
above described  (zones A, B, C, including at earlier times also zone
D) are known, one can calculate the X-ray luminosity from the whole 
remnant. This is given by
\begin{equation}
      \label{eq.x}
L_X = 4 \pi \int_0^{R_{out}} r^2 n^2 \Lambda_X(T,Z) {\rm d}r ,
\end{equation}
where $n(r)$ is the atomic number density, $R_{out}$ marks the
location of the outer shock and
$\Lambda_X(Z,T)$ is the X-ray emissivity derived by Raymond \& Smith 
(1977) in their hot-plasma code (see Strickland \& Stevens 2000).
Clearly, regions in which the gas temperature is below the X-ray 
cut-off temperature ($T_{cut} \approx 5 \times 10^5$K), would not 
contribute to the global X-ray emission.

\subsection{The X-ray luminosity from the star cluster and its free wind}

The temperature and density distributions within zones A and B, 
obtained from the numerical models (see Silich et al. 2004), lead to 
the corresponding X-ray luminosity. The results of the calculations
for a $10^6$\Msol \, clusters of 1 pc and 10 pc radii are presented in
Figure 2. The figure shows the contribution, 
$\epsilon_f = [L_{X,A+B} - L_X(R)] / L_{X,A+B}$, arising from
concentric shells with inner radius, $R$, and outer radius equal to the
X-ray cutoff radius, to the total X-ray luminosity, $L_{X,A+B}$, from
zones A and B. Figure 2 shows that the contribution from the free-wind 
region (zone B) is negligible. The value of $\epsilon_f$ for shells
with radius $R \ge R_{SC}$ never exceeds $\sim 20\%$ of the
cumulative emission from zones A and B. This is mainly due to the 
rapid decline in density ($\propto r^{-2}$) in the free wind region.
\begin{figure}[htbp]
\plotone{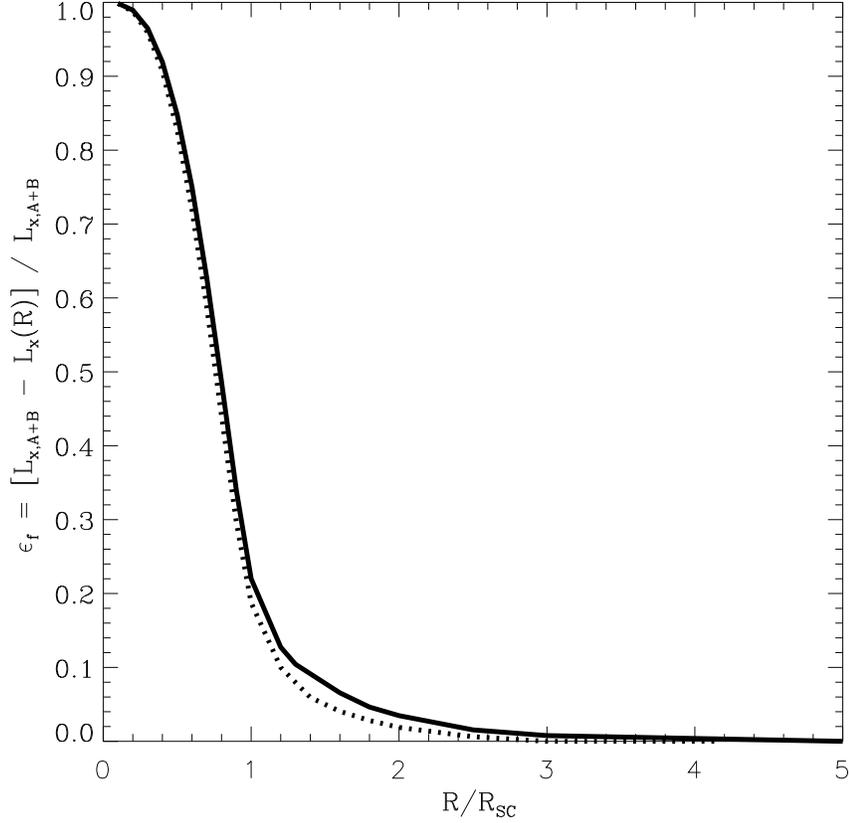}
\caption{The distribution of the radiated energy along the central
zones A and B. Solid and dotted lines display the results of the 
calculations for 
quasi-adiabatic (solid line) and strongly radiative (dotted line)
outflows. The star cluster mass and wind adiabatic
terminal speed are $10^6$\Msol \, and $V_{A,\infty} = 1000$ km s$^{-1}$
in both cases. The star cluster radii are $R_{SC} = 10$ pc (solid
line) and $R_{SC} = 1$ pc (dotted line). 
$L_{X,A+B} = 3.2 \times 10^{39}$ erg s$^{-1}$  and 
$L_{X,A+B} = 2.6 \times 10^{38}$ erg s$^{-1}$ in the case of 1 pc 
and 10 pc cluster, respectively.}
\label{fig2}
\end{figure}
A good estimate of $L_{X,A+B}$ can be obtained if one approximates the 
density and temperature distributions within zone A by constant values 
(see equations \ref{eq.1c}). The evaluation of integral (\ref{eq.x}) 
then yields
\begin{equation}
      \label{eq.6a}
L_{X,A+B} = \frac{4 \alpha^2_{\rho} \pi \Lambda_X \rho^2_{SC} 
             R^3_{SC}}{3 \mu^2_a} ,
\end{equation}
where $\mu_a  = 14/11 m_H$ is the mean mass per atom.
This simple analytic formula also accounts for the contribution
from zone B, and is in excellent agreement with the results of our 
semi-analytical calculations (see Figure 3) if the fiducial
coefficient, $\alpha_{\rho} \approx 2.0$. 
\begin{figure}[htbp]
\plotone{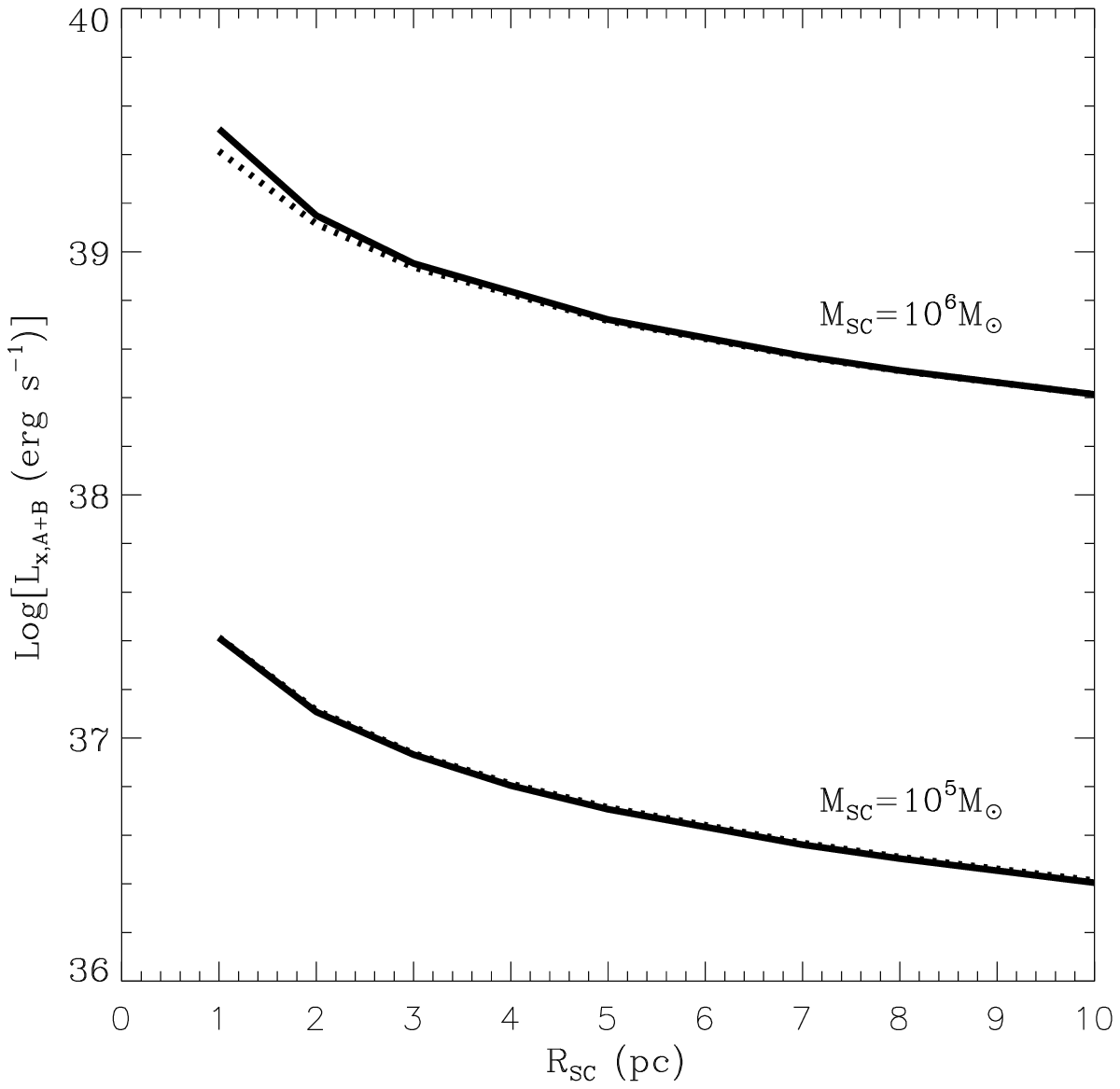}
\caption{The X-ray luminosity of SSCs and their winds (zones A+B) as 
a function of $R_{SC}$. The solid lines result from numerical 
calculations and are compared to the analytical formula (\ref{eq.4})
for star clusters with masses $10^6$\Msol \, and  $10^5$\Msol 
(dotted lines). The adiabatic wind terminal speed is 
$V_{A,\infty} = 1000$ km s$^{-1}$ in both cases, and the ejected gas 
metallicity is solar.}
\label{fig3}
\end{figure}
One can also obtain  $\rho_{SC}$ from the equation (\ref{eq.1c}) and 
re-write equation (\ref{eq.6a}) in the form
\begin{equation}
      \label{eq.4}
L_{X,A+B} = 3.8 \times 10^{34} \Phi(T_{SC},Z_{SC}) 
\frac{L^2_{38}}{R_{SC,1}V^6_{1000}} = 3.25 \times 10^{34} \Phi(T_{SC},Z_{SC}) 
\frac{L^2_{38}}{R_{SC,1}T^3_{keV}},
\end{equation}
where $T_{keV}$ is the temperature of the X-ray plasma measured in 
keV units. Equation (\ref{eq.4}) is normalized to 
$\Lambda_X = 3 \times 10^{-23}$ erg cm$^3$ s$^{-1}$ and the normalization 
function, $\Phi(T_{SC},Z_{SC})$, depends on the plasma temperature, 
$T_{SC}$, and its metallicity, $Z_{SC}$ (see Figure 4). It 
predicts a quadratic dependence of the X-ray luminosity (zones A and B)  
on the star cluster mass ($L_{SC}$ scales linearly with the mass of the 
cluster) and is similar to the scaling relation from Stevens \& 
Hartwell (2003; note a misprint in their equation 10 where a star 
cluster wind scaling parameter should be proportional to 
$V^{-2}_{A,\infty}$, see also Oskinova, 2005).
\begin{figure}[htbp]
\plotone{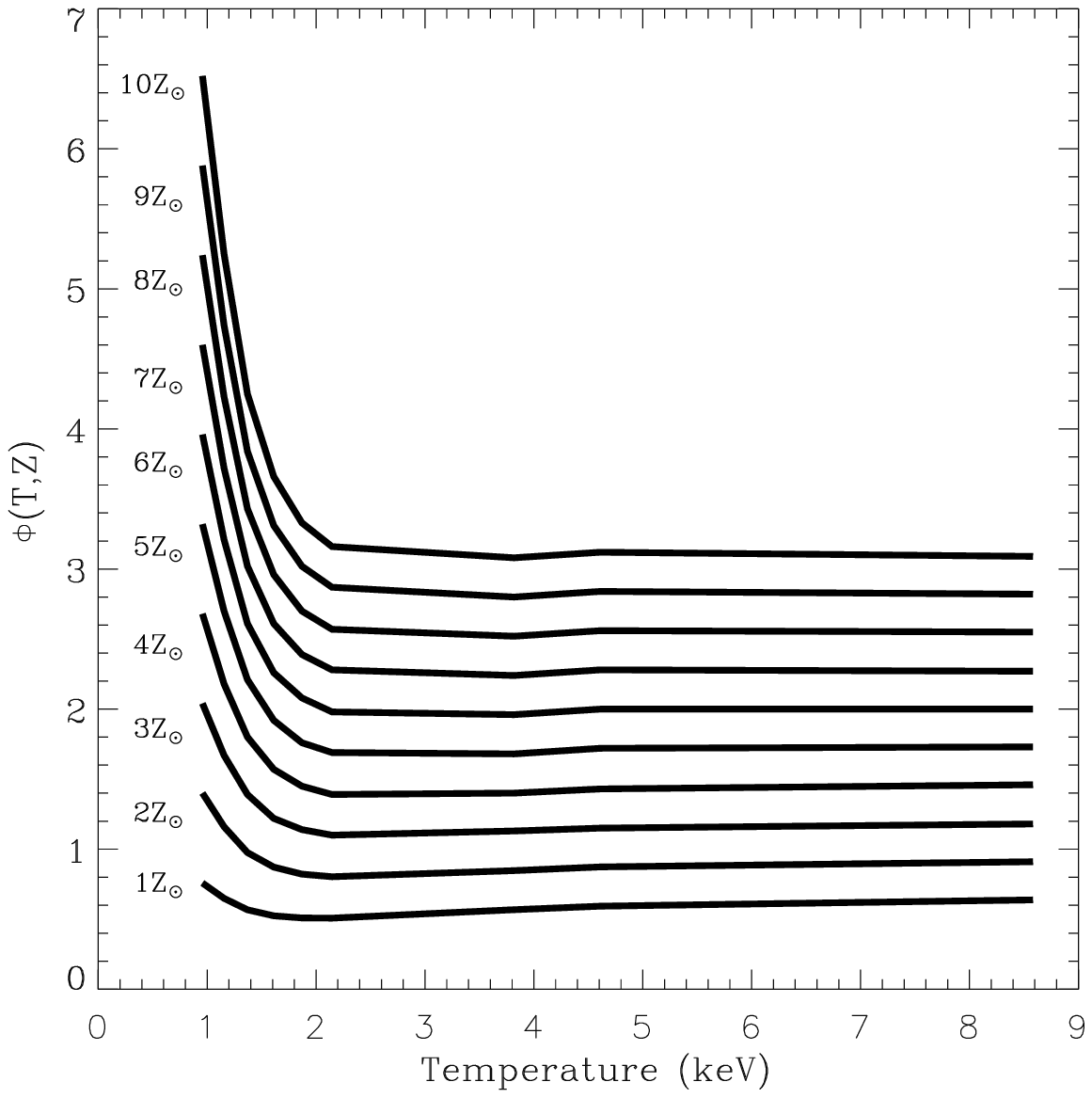}
\caption{The normalization function $\Phi(T,Z)$ plotted as function
of the X-ray plasma temperature measured in keV units. 
Different lines are marked with the assumed gas metallicity.}
\label{fig4}
\end{figure}

Equation (\ref{eq.4}) shows how the X-ray emission from the 
central zones A and B depends on the main star cluster parameters. 
The strong dependence on $V_{A,\infty}$
comes from the fact that close to the star cluster surface 
$\rho_w \sim {\dot M_{SC}} / R_{SC}^2 V_{A,\infty} \sim L_{SC}/V_{A,\infty}^3$.
This last relationship leads to equation (\ref{eq.4})
as the X-ray luminosity (along with radiative cooling) is
in direct proportion to the square of the gas number density. This implies
that the wind adiabatic terminal velocity (or the ratio of the energy to 
the mass deposition rates) is the key parameter that defines the
X-ray emission from the central zones. Note that equation (\ref{eq.4}) 
cannot be applied to very compact and massive star clusters above the 
threshold line (see Silich et al. 2004; Tenorio-Tagle et al. 2005),
clusters for which  the stationary wind solution is inhibited by 
radiative cooling.

\subsection{X-ray luminosity from the bubble}

\subsubsection{Adiabatic stage.}
During the early  stages, when $t < t_{cool}$, the distributions of density
and temperature in zones C and D are close to homogeneous and one
can evaluate integral (\ref{eq.x}) analytically: 
\begin{eqnarray}     
      & & \hspace{-1.0cm}  \nonumber
L_{X,C} = \frac{4 \pi}{3} \Lambda_X(T_C,Z_C) n^2_C 
          (R^3_{CD} - R^3_{in}) = 
           \frac{3 \Lambda_X(T_C,Z_C) L^2_{SC} 
           (t - \frac{R_{in}}{V_{A,\infty}})^2} 
           {\pi \mu^2_a V^4_{A,\infty} (R^3_{CD} - R^3_{in})} =  
      \\[0.2cm] \label{eq.7a}
      & & \hspace{-0.0cm}
9.5 \times 10^{34} \left(\frac{\Lambda_X}{3 \times 10^{-23}}\right)
\frac{n^{3/5}_{ISM} L^{7/5}_{38} [1 - R_{in}/V_{A,\infty} t]^2 t^{1/5}_7}  
{V^4_{1000}[1 - (R_{in}/R_{CD})^3]} \, erg \, s^{-1} ,
      \\[0.2cm] \nonumber
      & & \hspace{-1.0cm}
L_{X,D} =  \frac{4 \pi}{3} \left(\frac{\rho_{ISM}}{\mu_a}\right)^2
           \frac{\Lambda_X(T_D,Z_{ISM}) R^3_{out}}
           {1 - (R_{CD}/R_{out})^3} =
      \\[0.2cm] \label{eq.7b}
      & & \hspace{-0.0cm}
4.6 \times 10^{41} \left(\frac{\Lambda_X}{3 \times 10^{-23}}\right)
n^{7/5}_{ISM} L^{3/5}_{38} t^{9/5}_7 \, erg \, s^{-1} ,
\end{eqnarray}
where $n_C$ and $Z_C$ are the shocked wind (zone C) number density and 
metallicity, $T_C$ and $T_D$ are plasma temperatures in zones C and D, 
and $t_7$ is the evolutionary time measured in units of $10^7$ yr.
Thus, at the adiabatic stage the luminosity from the outer shell highly 
exceeds that from the hot bubble interior and dominates the 
total X-ray output. This is due to the much smaller plasma density in
zone C ($\rho_C/\rho_D \sim {\dot M_{SC}} t /
\rho_{ISM} R^3_{out} \sim V^2_{out}/V^2_{A,\infty} \ll 1$).
However, if the metallicity of the ISM is not well 
below the Solar value, this stage is so short (less than $10^5$ yr;  
see equation \ref{eq.3}), that we omit it in our further consideration.

\subsubsection{Hot bubble with a cold outer shell.}
At later times the swept up interstellar gas cools down to values well below 
the X-ray cut-off temperature and forms
a dense outer shell separated from the hot bubble interior by the
contact discontinuity (see Figure 1). After this stage, the X-ray 
luminosity of zone C  has been calculated by Chu et al. (1993). It is
\begin{equation}
      \label{eq.5}
L_{X,C} = 3.4 \times 10^{36} Z_{C} I(\tau) L_{38}^{33/35} n_{ISM}^{17/35}
               t_7^{19/35} erg \, s^{-1} ,
\end{equation}
where $Z_C$ is the metallicity of the X-ray plasma in zone C,
I$(\tau)$ is a dimensionless integral with a value close to 2. 
In the models of Chu et al. (1993)  the main 
contribution to the bubble X-ray luminosity comes from the outer, 
densest layers of zone C, where the density and temperature are 
determined by thermal evaporation of the cold outer shell.
For this reason the metallicity of the X-ray gas in this zone 
may be different from  that in zones A and B and hardly ever exceeds 
a few times the solar value (see Silich et al. 2001). Hereafter we assume 
that the chemical composition of the plasma in zone C is solar, 
$Z_C = Z_{\odot}$. On the other hand the temperature decreases towards
the outer layers (see equation \ref{eq.5a}) and therefore the X-ray emission
from zone C should be much softer than that from zones A and B.

\section{Integrated properties of the model predicted X-ray emission 
         and comparison with observations}

\subsection{Wind vs bubble luminosity}

We first used our analytic results (equation \ref{eq.4}) to derive
the expected X-ray luminosity from clusters of different masses
and different radii. The calculated X-ray luminosities
from zones A and B for a range  of the star cluster wind terminal speeds
(1000 km s$^{-1} \le V_{A,\infty} \le 1500$ km s$^{-1}$)
are presented in Figure 5. The expected temperatures of the plasma
are $\sim 1.1 \times 10^7$K and $\sim 2.5 \times 10^7$K for 
$V_{A,\infty} = 1000$ km s$^{-1}$ and $V_{A,\infty} = 1500$ km s$^{-1}$, 
respectively (see equation \ref{eq.1a}). The normalization functions are 
$\Phi = 0.758$ and $\Phi = 6.52$ for $T \sim 1.1 \times 10^7$K plasma
and $\Phi = 0.508$ and $\Phi = 3.16$ for $T \sim 2.5 \times 10^7$K for
a plasma with $Z = Z_{\odot}$ and $Z = 10Z_{\odot}$, respectively 
(see Figure 4). Figure 5 shows that
the X-ray luminosity from the cluster (zones A and B) is strongly
dependent on the star cluster parameters and may easily vary within 
several orders of magnitude (the low luminosity limits move towards 
smaller values if $V_{A,\infty}$ exceeds 1500 km s$^{-1}$) even for a 
sample of clusters with a given mass. The crucial parameter, and 
unfortunately not a well known parameter, is the star cluster wind 
terminal speed, $V_{A,\infty}=(2 L_{SC}/{\dot M}_{SC})^{1/2}$. 
Perhaps the only reasonable way to have observational
restrictions for this parameter for distant clusters is to measure 
the temperature of the X-ray plasma to then derive $V_{A,\infty}$ 
from equation (\ref{eq.1a}). This possibility may provide reasonable
estimates for the most massive and compact clusters despite the
presence of X-ray binaries (see below).

The luminosities are also dependent on
the metallicity of the X-ray plasma (compare panels a and b in
Figure 5). For more metallic plasmas the calculated luminosities
are shifted to higher values and the difference between the
low and the high velocity (temperature) outflows becomes even more
significant as the enhanced metallicity affects less the
emissivity of the higher temperature plasma (see Figure 4). 
\begin{figure}[htbp]
\plottwo{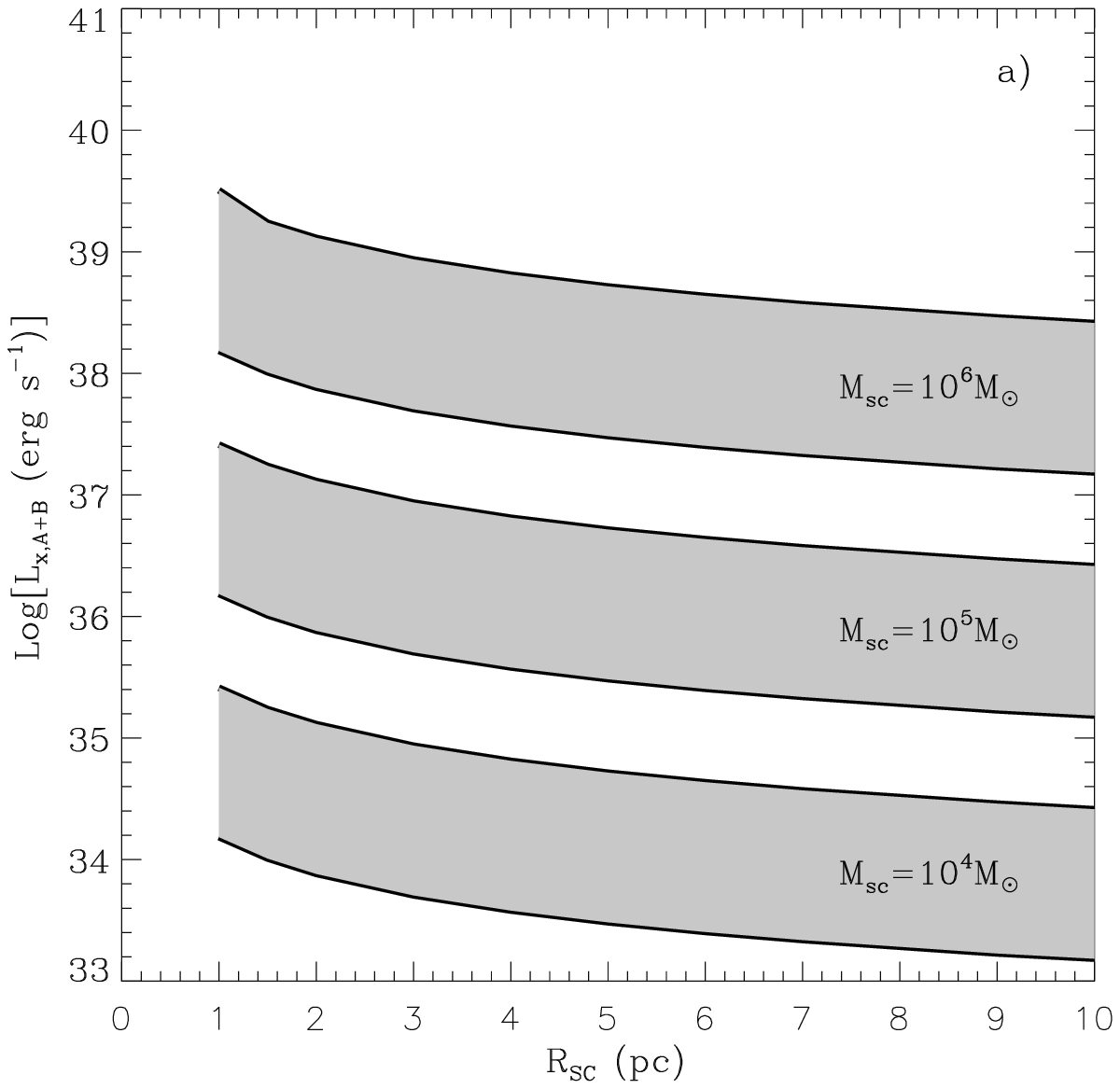}{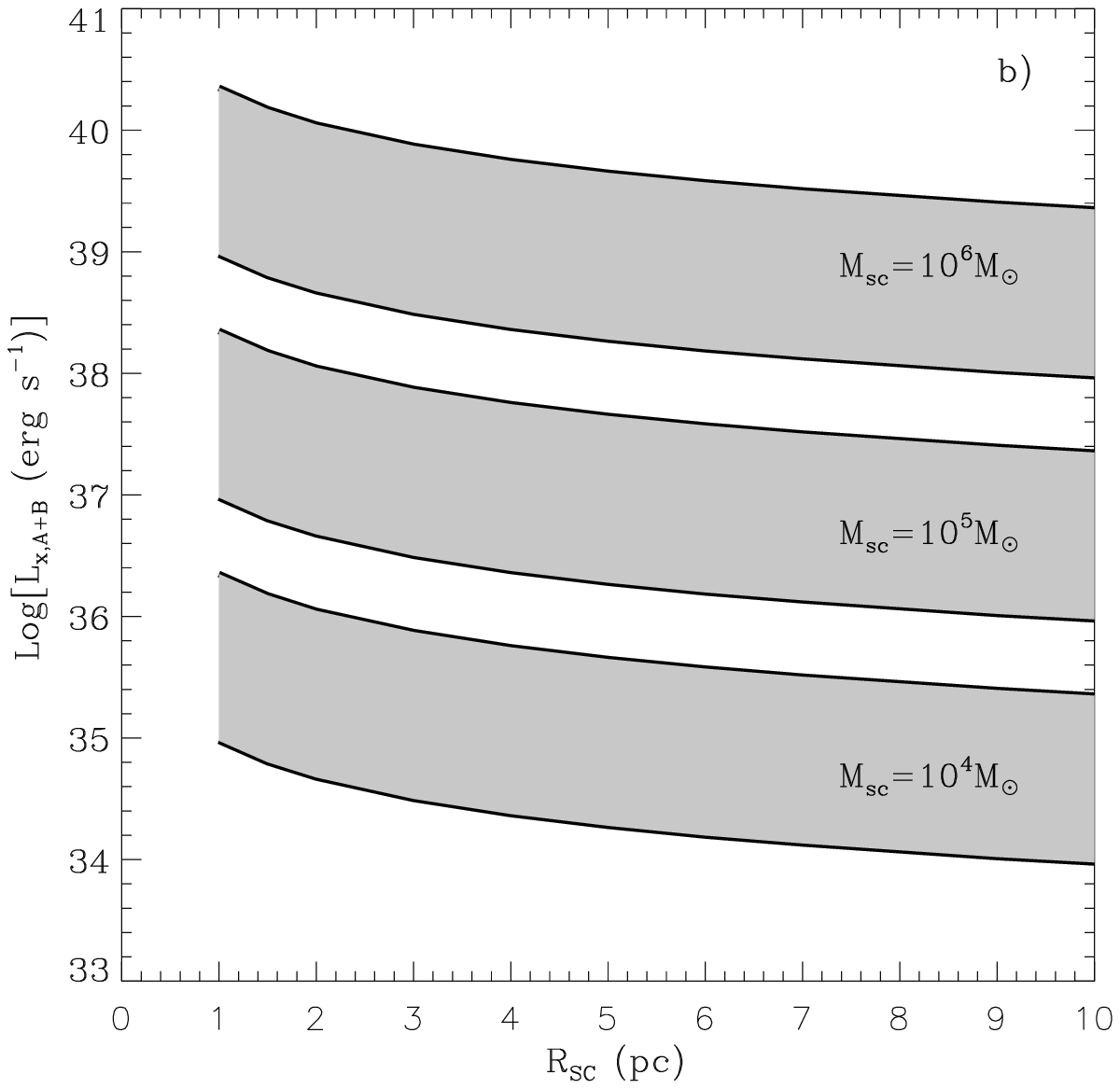}
\caption{The X-ray luminosity of SSCs and their winds (zones A+B) as 
a function of star cluster parameters. The shadow regions on the
diagram represent (from the top to the
bottom), the calculated X-ray luminosity for $10^6$\Msol ,
$10^5$\Msol \, and $10^4$\Msol \, clusters, respectively. 
$V_{A,\infty}=1000$ km s$^{-1}$ and 
$V_{A,\infty}=1500$ km s$^{-1}$ were used to calculate the upper and
lower luminosity limits for every star cluster. Panel a displays
the results of the calculations for a plasma with solar metallicity,
and panel b is the same for a plasma with Z=10Z$_{\odot}$.}
\label{fig5}
\end{figure}

We have also compared the X-ray emission from zones A and B 
with the X-ray emission from the hot bubble interior (zone C) predicted 
by the Chu et al. (1993) standard model. 
\begin{figure}[htbp]
\vspace{19.5cm}
\includegraphics{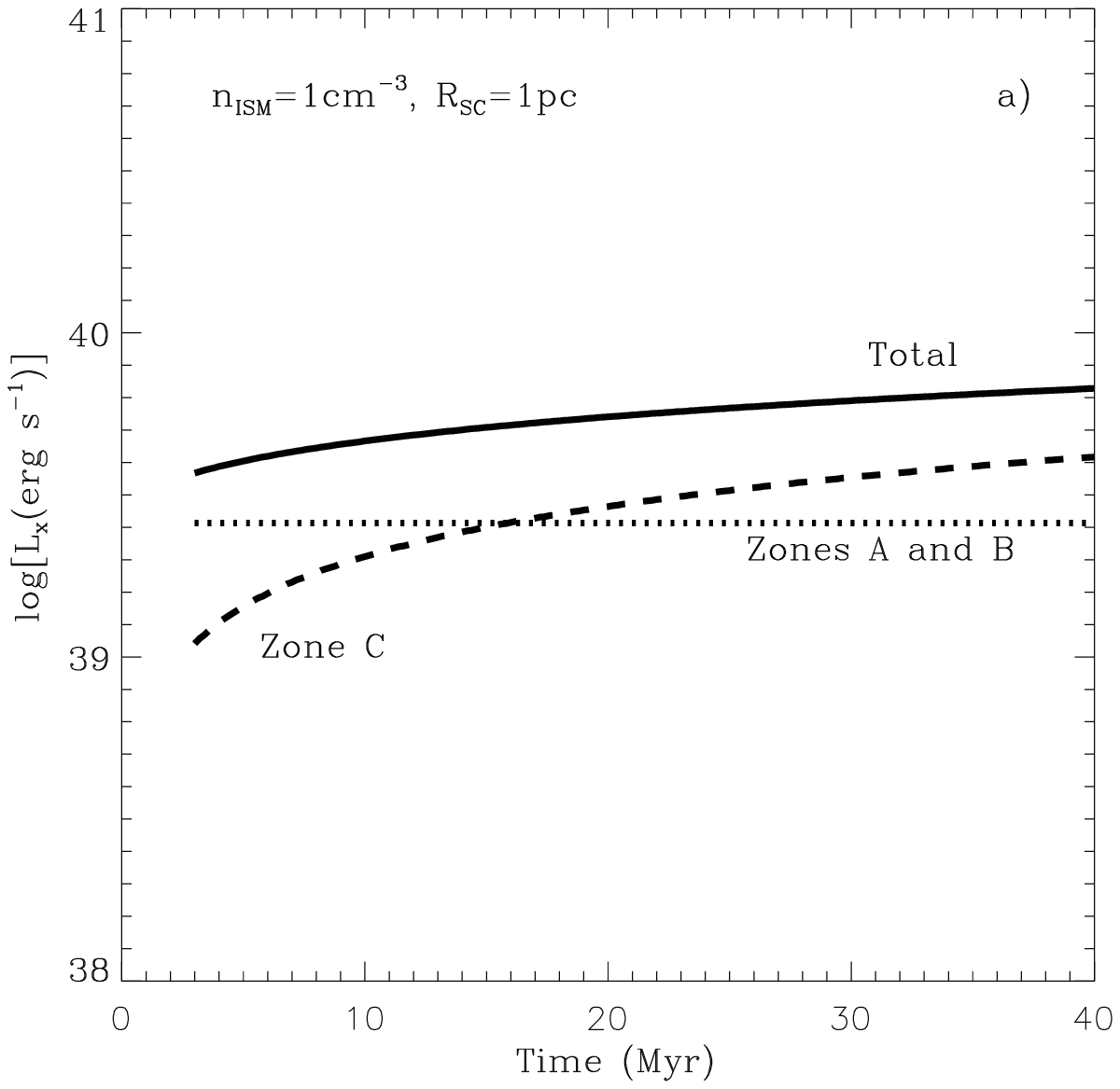}
\includegraphics{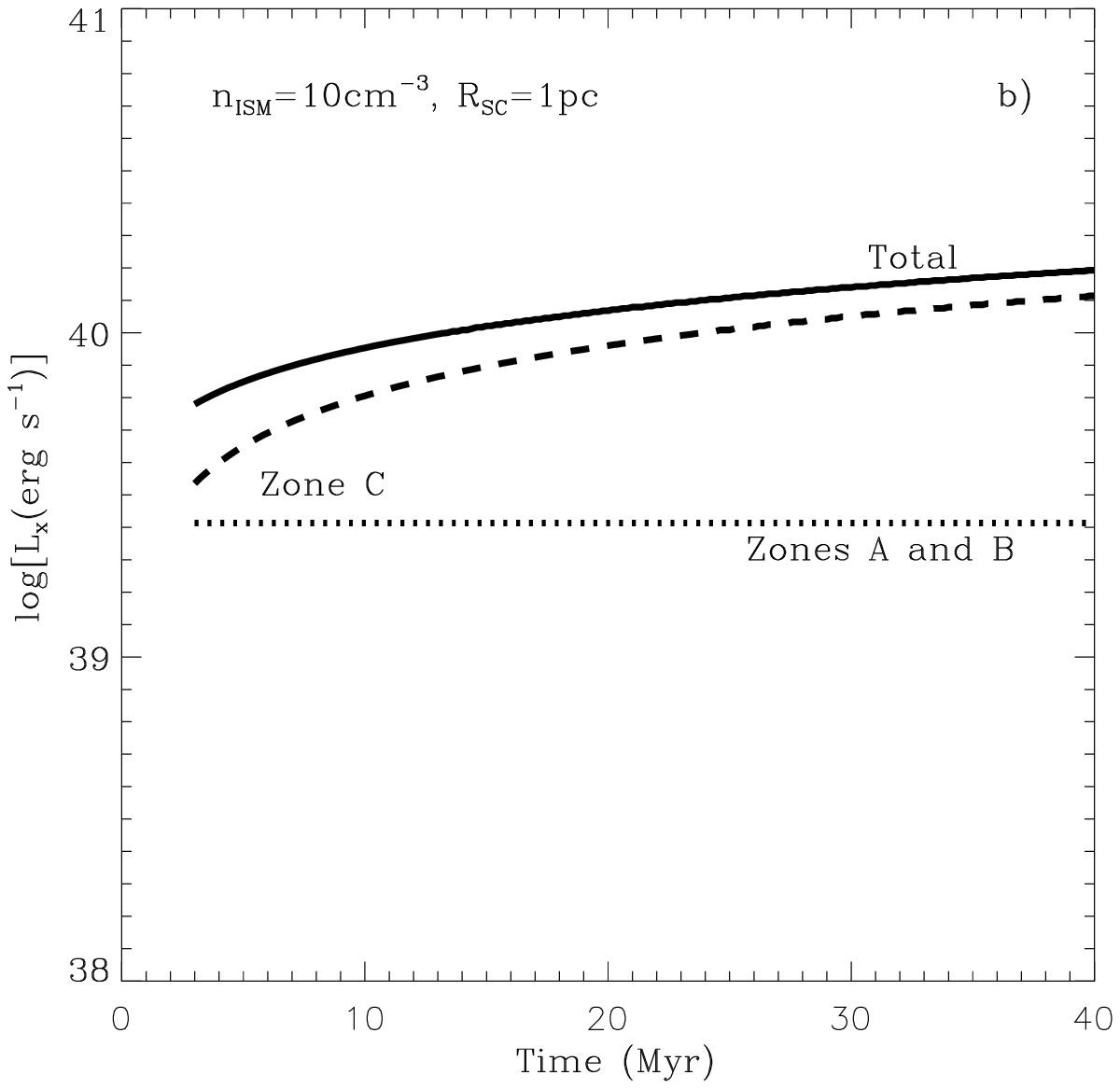}
\includegraphics{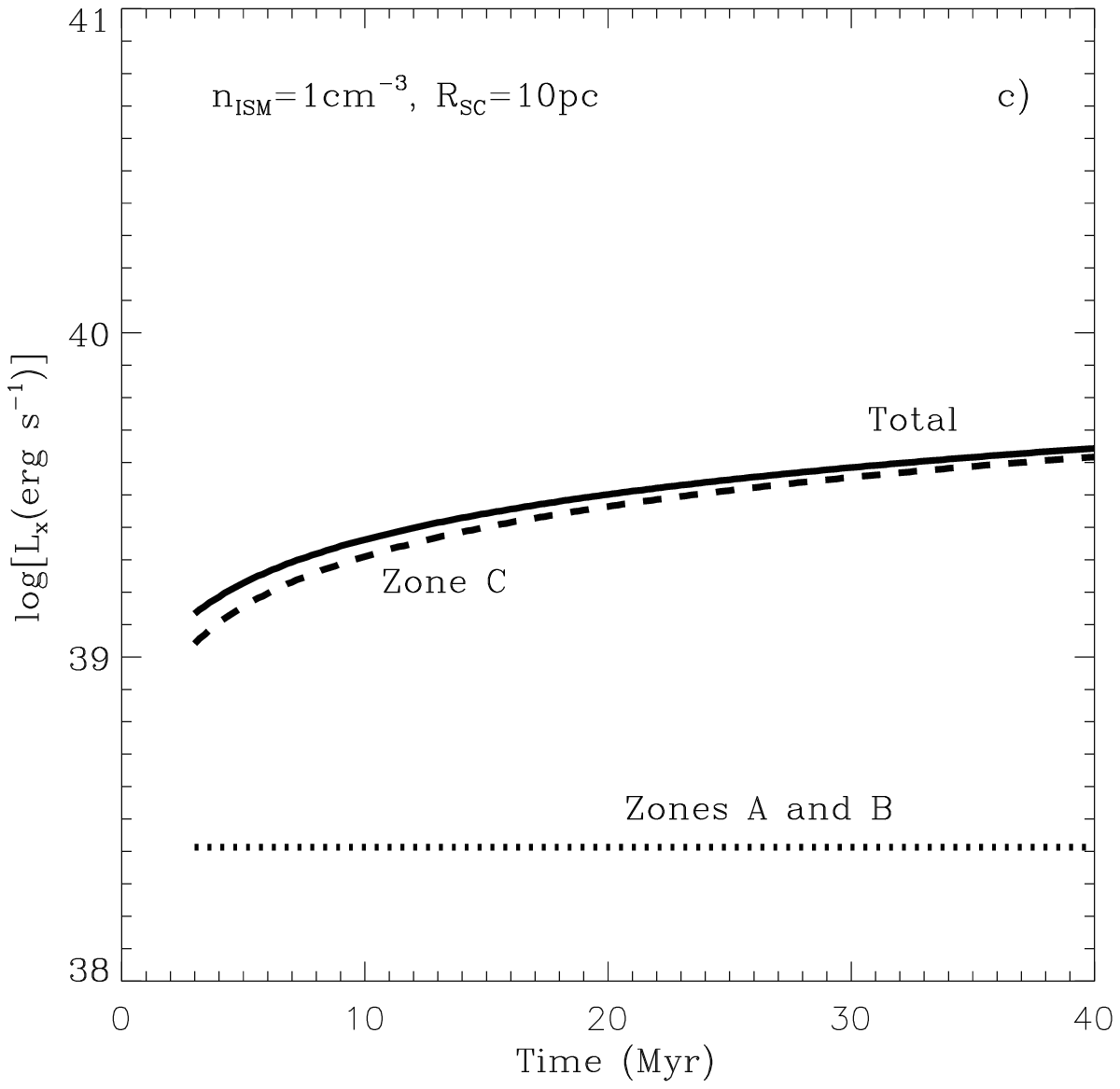}
\includegraphics{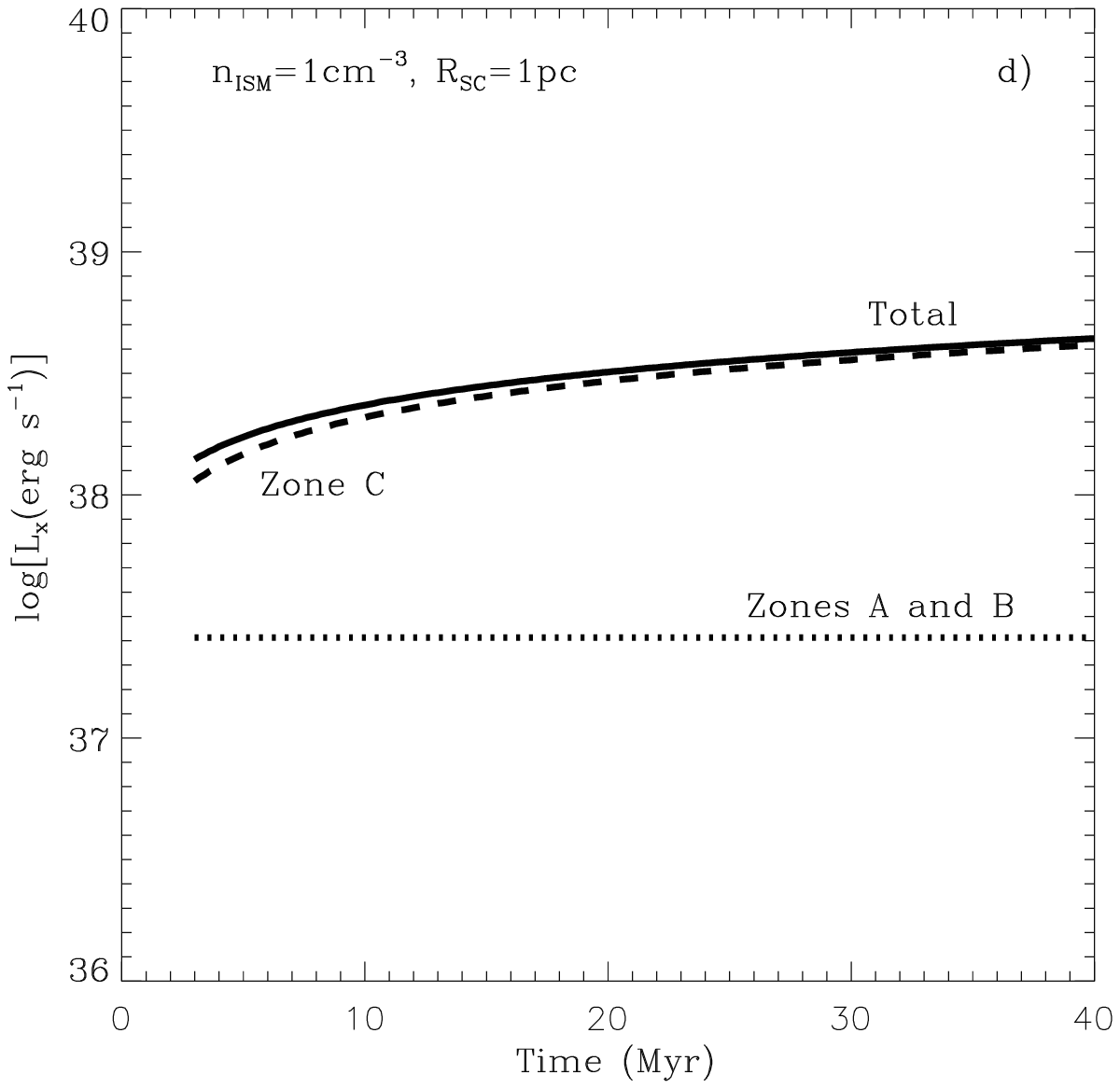}
\includegraphics{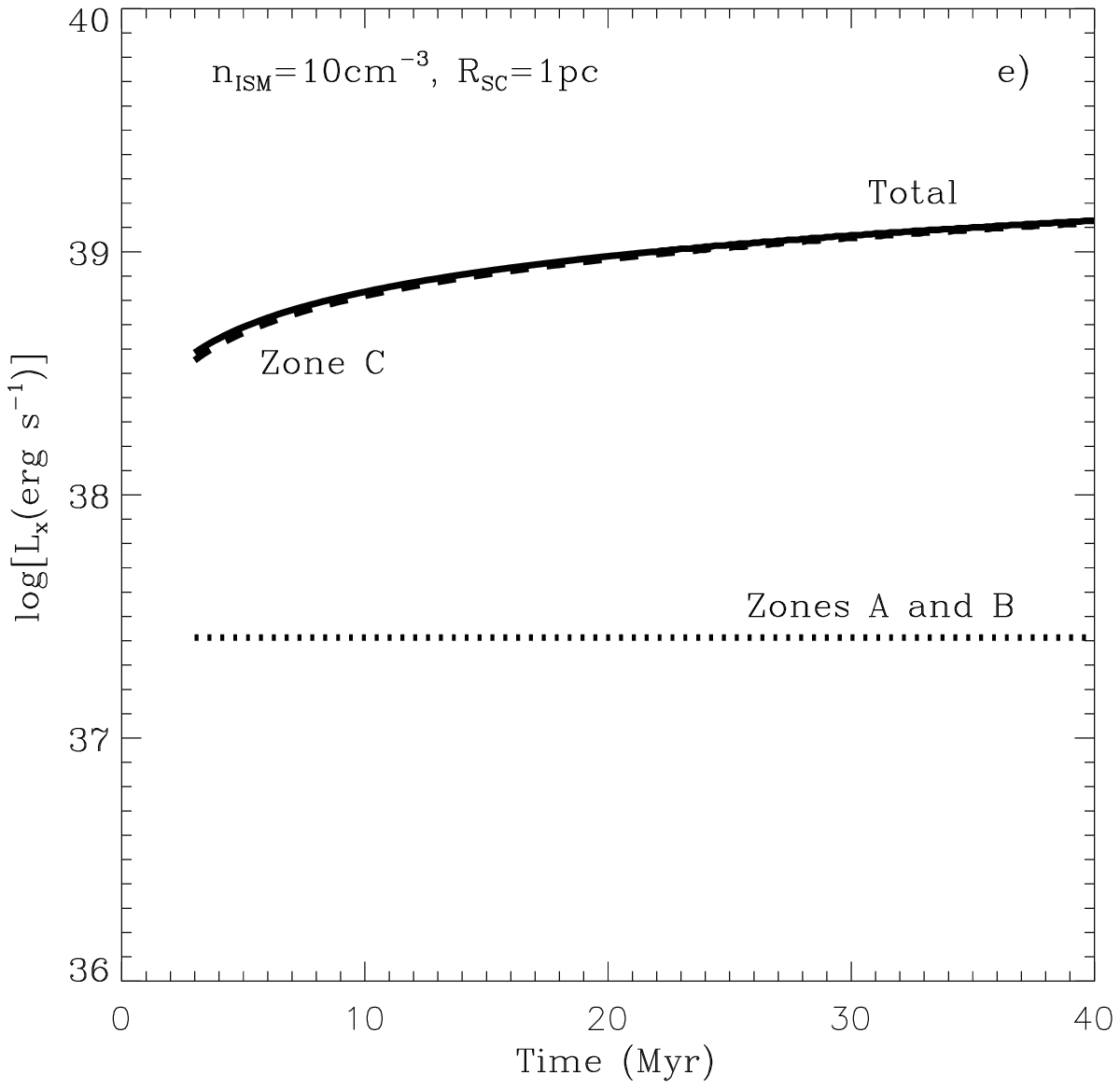}
\includegraphics{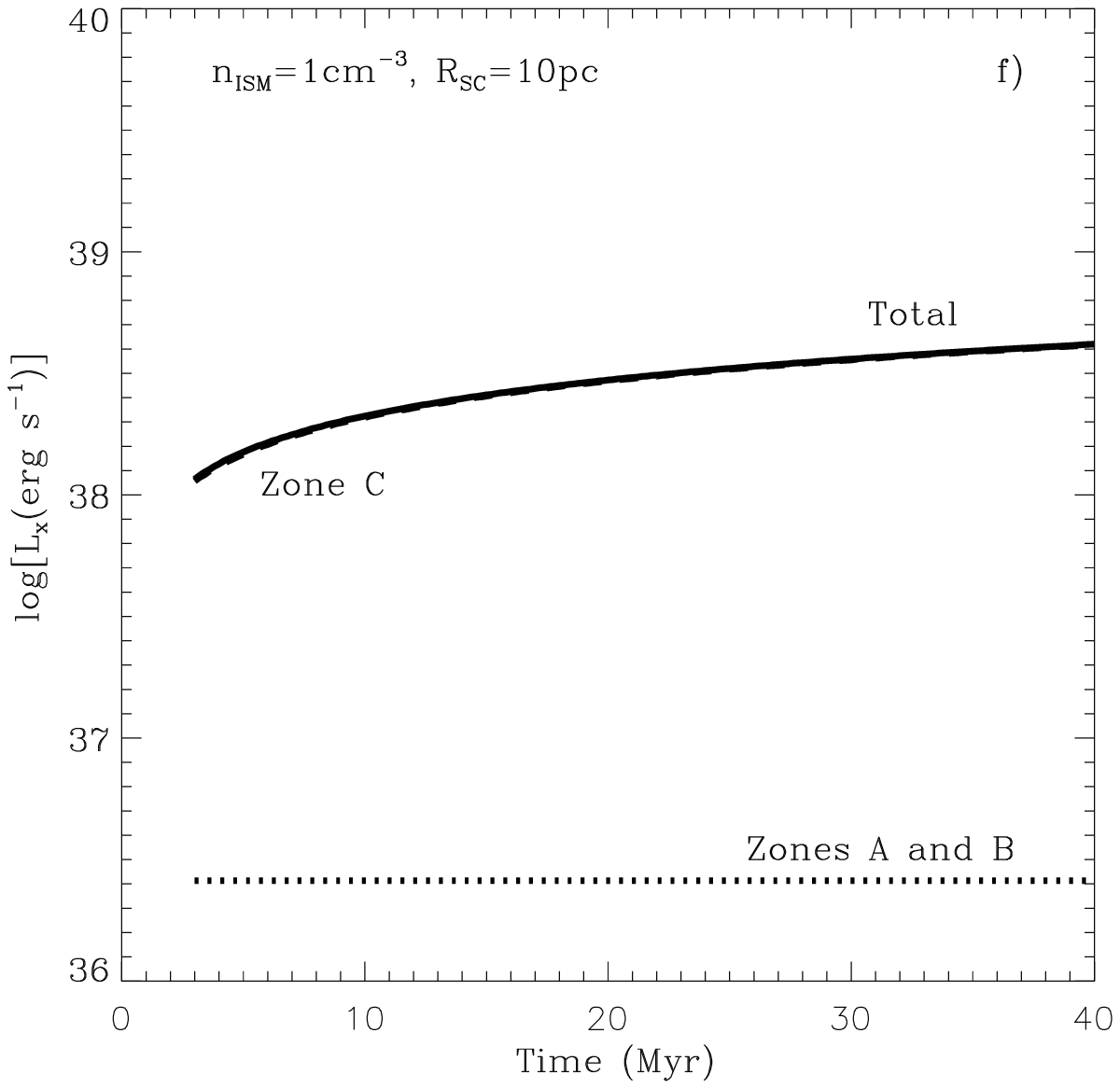}
\caption{X-ray luminosity generated by star clusters of different masses
and radii located in different ISM environments. 
a) $10^6$\Msol \, 1 pc cluster in the low density ISM with 
$n_{ISM} = 1$ cm$^{-3}$; 
b) $10^6$\Msol \, 1 pc cluster in the dense ISM with 
$n_{ISM} = 10$ cm$^{-3}$; 
c) $10^6$\Msol \, 10 pc cluster in the low density ISM with 
$n_{ISM} = 1$ cm$^{-3}$;
d) $10^5$\Msol \, 1 pc cluster in the low density ISM with  
$n_{ISM} = 1$ cm$^{-3}$;
e) $10^5$\Msol \, 1 pc cluster in the dense ISM with  
$n_{ISM} = 10$ cm$^{-3}$; 
f) $10^5$\Msol \, 10 pc cluster in the low density ISM with  
$n_{ISM} = 1$ cm$^{-3}$; 
Solid lines display the total X-ray luminosity. Dotted lines
are contribution from zones A and B, and dashed lines
are contribution from zone C, respectively.}
\label{fig6}
\end{figure}
Figure 6 shows the contributions from the star cluster (zones A and B) 
and from the expanding superbubble (zone C) to the total 0.3 keV - 8.0 keV 
emission, for clusters of different masses and radii evolving into an
ISM with different densities. The X-ray emission from the high
temperature zones A and B may dominate only in the case of very
massive ($\ge 10^6$\Msol) and compact (few parsecs) star clusters 
evolving into a low density ISM (panel a in Figure 6). Otherwise 
the total luminosity is dominated by the hot superbubble interior. 
The contribution from the bubble plasma progressively increases
with time and becomes dominant after a short while even for 
compact and massive clusters, if they evolve into a dense ISM (see panel b, 
Figure 6). The X-ray luminosity from star clusters 
decreases for larger clusters (see equation \ref{eq.4}) 
shifting the luminosity from zones A and B well below the bubble
luminosity even for massive clusters evolving into a low
density ISM (see panel c in Figure 6). Clearly, this tendency
becomes even stronger for less massive clusters (panels d, e, f
in Figure 6). However, even in the case of the rather dense ISM
($n_{ISM} = 10$ cm$^{-3}$), the radius of the bubble exceeds 90 pc
after 1 Myr, in the case of a $10^5$\Msol \, cluster and 150 pc in 
the case of a $10^6$\Msol \, cluster.
This implies that compact and massive star clusters are to
be detected as point-like X-ray sources embedded into extended 
regions of diffuse X-ray emission. 

One can also distinguish between zones A and B and zone C from 
their contributions to the hard and soft energy bands. Indeed,
the temperature of the hot plasma ejected from the star cluster
region (zone A) is defined by the ratio of the mechanical energy
and mass deposition rates (or by the terminal speed of the star
cluster wind) and, for reasonable values of $V_{A,\infty}$, falls
in the range 1 keV - 10 keV (see equation \ref{eq.1a}). At the
same time the temperature of the plasma which dominates the bubble X-ray
emission is much lower due to dilution with the cold material
evaporated from the outer shell. This suggests that the soft component
to the X-ray emission is dominated by the bubble plasma whereas
the hard component is defined by the hot thermalized ejecta
(zone A and B). This conclusion is illustrated by Figure 7 which
presents the results of the numerical calculations for a $10^5$\Msol \,
cluster of 1 pc radius evolving into a 1 cm$^{-3}$ ISM (shown also 
in Figure 6d). Figure 7 clearly demonstrates that the star cluster 
ejecta dominates the hard X-ray emission even if the total luminosity 
is completely defined by the bubble plasma. Note that because
of the assumed constant emissivity of the X-ray plasma, formula
(\ref{eq.5}) overestimates the X-ray emission from the bubble
interior by a factor $\sim 2$. 
\begin{figure}[htbp]
\plottwo{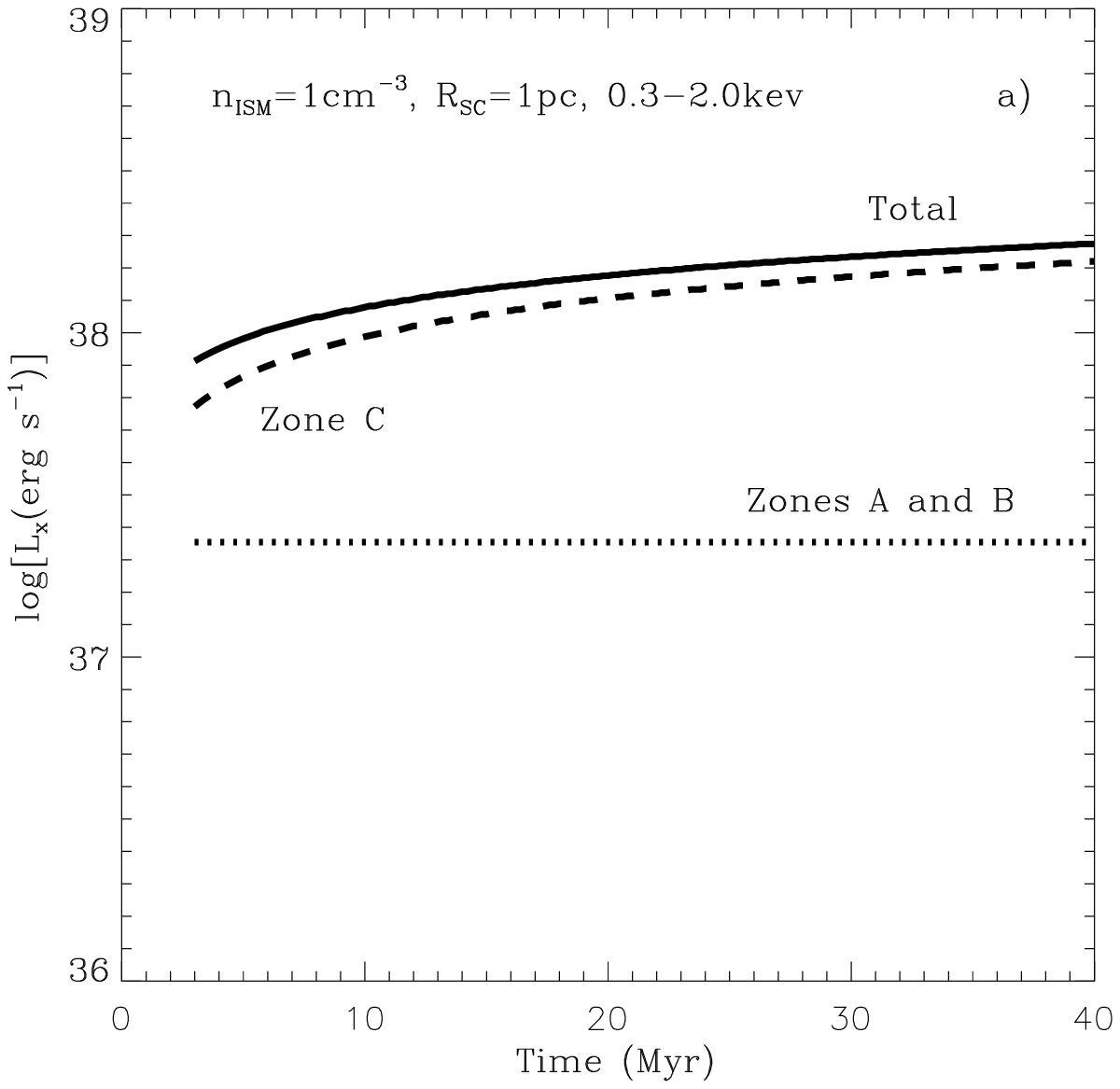}{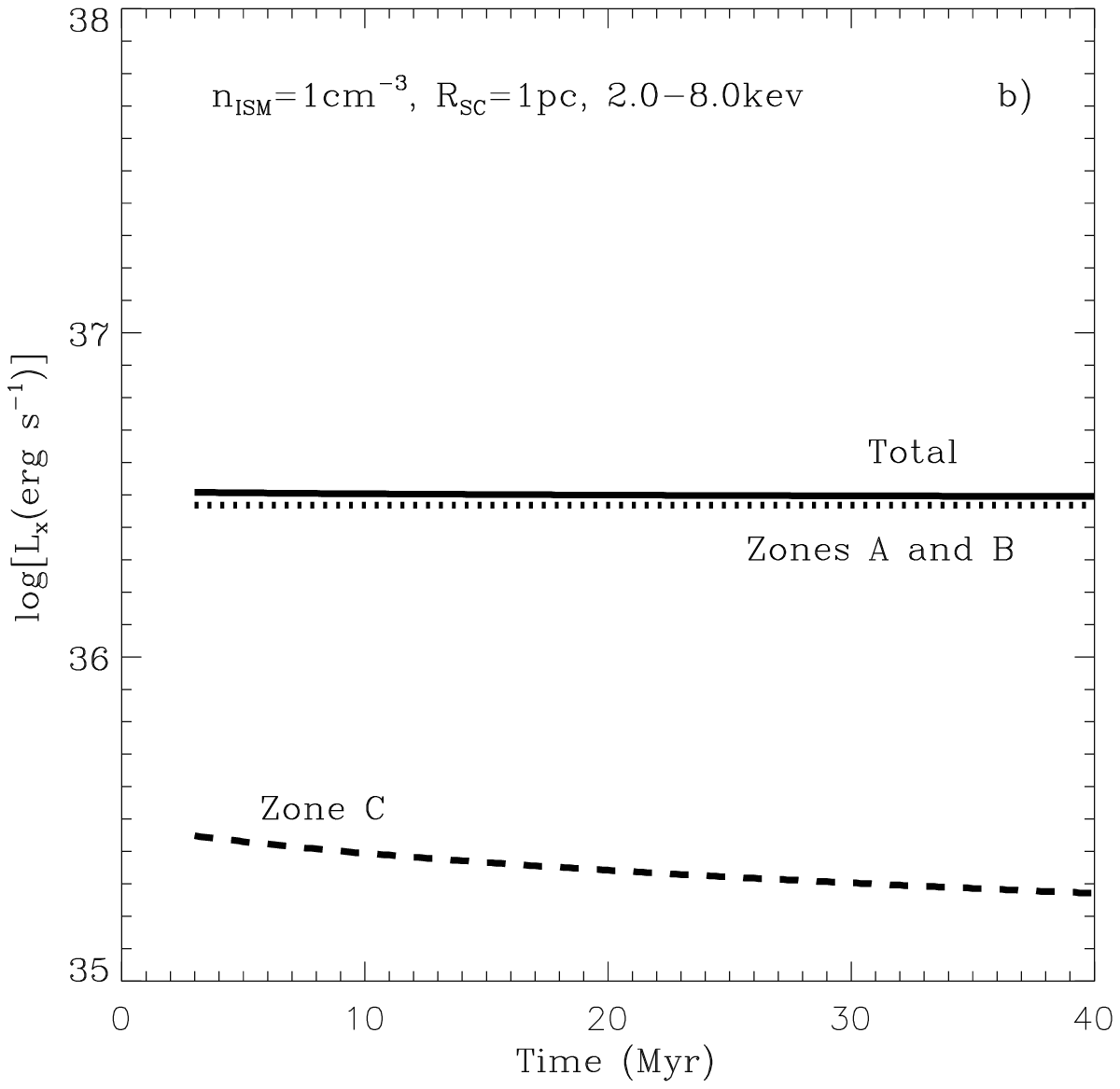}
\caption{Soft and hard components of the X-ray luminosity 
generated by the SSC. The contribution of different zones
to the soft (0.3 keV -2.0 keV) and hard (2.0 keV - 8.0 keV) 
X-ray emission is displayed in panels a and b, respectively. 
The star cluster mass is $10^5$\Msol , radius 1 pc, and the
star cluster wind terminal speed is 1000 km s$^{-3}$. The density
of the ISM is 1 cm$^{-3}$. The metallicity was assumed to be 
Z=Z$_{\odot}$ in all zones.} 
\label{fig7}
\end{figure}

From equation (\ref{eq.4}) one can derive the fraction, $\epsilon_X$, 
of the injected mechanical energy, $L_{SC}$, that is transformed into 
the X-rays. The X-ray production efficiency, $\epsilon_{A+B}$, for 
the hard component associated with the star cluster plasma is  
\begin{equation}
      \label{eq.8}
\epsilon_{A+B} = \frac{L_{X,A+B}}{L_{SC}} =
             3.8 \times 10^{-4} \Phi(T_{SC},Z_{SC}) 
             \frac{L_{38}}{R_{SC,1}V^6_{1000}} .
\end{equation}
For $10^6$\Msol \, clusters with $L_{38} = 3 \times 10^2$ it hardly 
exceeds 10\% even for most compact clusters and decreases
progressively for star clusters with smaller masses.
 
One can obtain the X-ray production efficiency, $\epsilon_C$, for the
soft bubble plasma from equation (\ref{eq.5}): 
\begin{equation}
      \label{eq.9}
\epsilon_C = 3.4 \times 10^{-2} Z_{C} I(\tau) L_{38}^{-2/35} n_{ISM}^{17/35}
               t_7^{19/35} ,
\end{equation}
It is weakly dependent on the star cluster parameters, but is dependent on
the density of the ISM and becomes larger as the evolutionary time grows.
$\epsilon_C$ was discussed by Cervi\~no et al. (2002) who finally 
chose an arbitrary value of $\epsilon_C = 20\%$, and by Smith
et al. (2005) who collected published data for well-studied HII regions
and superbubbles and concluded that for young objects $\epsilon_C
\approx 0.02\%$, while for older clusters $\epsilon_C$ falls into the
range 0.2\% - 7\%.

\subsection{Comparison with nearby clusters}

A comparison of the model predictions with observed X-ray luminosities
requires that parameters of the embedded star cluster (half-light
radius, $R_{SC}$, mass, $M_{SC}$ and age, $\tau_{SC}$) will
be obtained from the optical or infrared observations and then be
compared with the observed X-ray luminosity, temperature (which is
proportional to the star cluster wind terminal speed, $V_{A,\infty}$)
and metallicity of the X-ray plasma. The full set of required
parameters are available only for a restricted sample of nearby
clusters. Below we confront our model (equation \ref{eq.4}) with
several nearby young stellar clusters. In contrast with Stevens \& 
Hartwell (2003), we will use for the comparison an observed
temperature of the X-ray plasma which is an appropriate parameter for
distant clusters, and will avoid to use the mean weighted terminal 
speed derived from the analysis of the individual massive stars, 
embedded members of the  cluster.

\subsubsection{The sample of clusters}

\noindent{\bf A) NGC 3603.} This cluster appears to be one of the
densest and more massive cluster known in our Galaxy. Within a 1 pc radius
NGC 3603 reveals a remarkable similarity in the central density and
stellar density distribution of R136, the dense core of 30 Dor in the
Large Magellanic Cloud (Moffat et al. 1994). However outside
$R_{SC} \approx 1$~pc, the stellar density in NGC 3603 decreases sharply
whereas the stellar density in 30 Dor remains to decrease with a
similar slope to larger radii. Malumuth \& Heap (1994) measured the
ionizing flux from 30 Dor as a function of distance from the cluster
center. They found that the number of ionizing photons from the inner
1 pc ($4^{\prime\prime}$) region is N$^0 \approx 1.7 \times 10^{51}$
s$^{-1}$ (see their figure 16). This corresponds to a $1.7 \times
10^4$\Msol \, cluster if compared with Leitherer \& Heckman (1995) 
evolutionary synthesis model and approximately two times smaller than the 
$\sim 3 \times 10^4$\Msol \, value obtained by  Brandl et al. (1996)
for the inner $20^{\prime\prime}$ region of 30 Dor. 
Because of the identical mass distribution in R136 and NGC 3603 inside
the inner 1 pc radius, we adopt $1.7 \times 10^4$\Msol \, value as a
total mass of the NGC 3603 cluster. Sung \& Bessel (2004) used  
optical color-magnitude diagrams for massive members of NGC 3603
and concluded that the age of the cluster is $\tau_{SC} = 1 \pm 1$~Myr.
The X-ray emission from the cluster has been studied by
Moffat et al. (2002) who detected around 40 point-like X-ray sources
definitely associated with the star cluster and a local diffuse
emission most probably associated with the hot plasma inside the star
cluster and with the free wind region. Moffat et al. (2002) concluded that 
approximately 20\% ($2 \times 10^{34}$ erg s$^{-1}$) of the total 
$1 \times 10^{35}$ erg s$^{-1}$ X-ray luminosity in the $0.5 - 10$ keV
band is associated with the local diffuse component of solar abundance 
and the best fitted temperature of $k T_{core} \approx 3.1$ keV at the 
core and $k T \approx 2.1$ keV (which is close to the
temperature at the star cluster edge that is expected from the star 
cluster wind model) in the outer region. 

\noindent{\bf B) Arches cluster.} The Arches cluster is another young
($\tau_{SC} = 2 \pm 1$~Myr; Figer et al. 1999a) and very compact
($R_{SC} \approx 0.2$~pc; Figer et al. 1999b) Galactic cluster. The
mass of the cluster has been estimated by Figer et al. (1999a) who
extrapolated the observed number of massive stars down to the low
cutoff mass limit. If a 1\Msol \, low mass cutoff is adopted, the star
cluster mass would be $M_{SC} \approx 1.1 \times 10^4$\Msol. The X-ray
luminosity from the cluster was first detected by Yusef-Zadeh et
al. (2002) who revealed within the volume occupied by the star cluster
two point-like sources embedded into a more extended (down to a 1.2-1.8
pc radius) X-ray halo. Law \& Yusef-Zadeh (2004) found that the
observed X-ray emission may be equally well fitted with an
overabundant, $Z = (4-5) Z_{\odot}$, one or two temperature thermal 
model, or with the one temperature plus power-law model with 
$\sim 1/6$ of the flux attributed to the power-law component. 
The required temperature of the thermal plasma is around
$k T = 1.5$~keV. The models give $L_X = (0.5 - 1) \times 10^{35}$ erg
s$^{-1}$ for the 0.5 - 8.0 keV energy band. Overall, the point-like
sources inside the cluster contribute $\sim 60$\% to the total
emission and the rest is distributed throughout the region with
dimensions of approximately $3.6 \times 2.4$ pc. This leads to
the X-ray luminosity from the cluster plasma, $L_X = (2 - 4) \times
10^{34}$ erg s$^{-1}$, that is slightly larger than that from the 
earlier estimates of Yusef-Zadeh et al. (2002; $L_X = 1.6 \times 
10^{34}$ erg s$^{-1}$). However the interpretation of the observed
X-ray emission is ambiguous. The observed spectrum may also be fitted
by the combined power law plus a 6.4 keV Gaussian component that
models a fluorescent Fe K$_{\alpha}$ emission from a nearby
molecular cloud. If this is the case, the contribution of the
fluorescent component may confuse the estimated X-ray emission
from the cluster plasma. 

\noindent{\bf C) Quintuplet cluster.} The Quintuplet cluster is the
smallest and the oldest ($\tau_{SC} = 4 \pm 1$~Myr, Figer et
al. 1999b) one. The mass of the cluster is $M_{SC} = 8.8 \times
10^3$\Msol \, if a 1\Msol \, low mass cutoff is used (Figer et al., 1999a),
and the radius is about of 1 pc (Figer et al. 1999b). The
X-ray emission from the cluster was detected by Chandra. After
subtraction of the four point-like sources, the X-ray spectrum of 
the Quintuplet cluster is perfectly fitted by the thermal plasma model
with the plasma temperature 2.4 keV at the star cluster center and 
a solar metallicity (Law \& Yusef-Zadeh, 2004). The
absorption-corrected halo emission in the $0.5 - 8$ keV energy band 
is $1 \times 10^{34}$ erg s$^{-1}$ (Law \& Yusef-Zadeh, 2004). From 
the observed central temperature we derive a plasma temperature at 
the star cluster surface (see section 3), $k T = 0.75 \times 
2.4 = 1.8$ keV, and then use it as an input parameter for our 
model.

\subsubsection{The comparison with model predictions}

In order to compare the observed X-ray luminosities with that
predicted by the star cluster wind model, we take an advantage of our
equation (\ref{eq.4}) and combine it with input parameters
obtained from the optical examination of the host clusters. 
To calculate the energy deposition rate, $L_{SC}$, provided by
multiple stellar winds and supernovae explosions inside the cluster,
we used a starburst synthetic model of Leitherer \& Heckman (1995)
and scale it with the star cluster mass. For a standard, $10^6$\Msol,
cluster we adopt a $10^{40}$ erg s$^{-1}$ energy deposition rate
before the first supernova explosion ($t \le 3$~Myr), 
and approximately three times larger average value after the 
beginning of the supernovae era. The star clusters parameters,
the observed and the calculated X-ray luminosities are collected
in Table 1.
\begin{table}
    \label{tab1} 
      \caption{Star clusters parameters and X-ray luminosities}
       \small
        \begin{flushleft}
\begin{tabular}{|l|c|c|c|c|c|c|c|c|} \hline 
Cluster & $M_{SC}$ & $R_{SC}$ & Age & $L_{SC}$ & T$_{obs}$ & Z & $L_{X, obs}$
& $L_{X, mod}$  \\
& (\Msol) & (pc) & (Myr) & (erg s$^{-1}$) & {keV} & (Z$_{\odot}$) & 
(erg s$^{-1}$) & (erg s$^{-1}$) \\
\hline
NGC 3603 & $1.7 \times 10^4$ & 1 & 1 $\pm 1$ & $1.7 \times 10^{38}$ & $2.1 \pm
0.7$ & 1 & $2.0 \times 10^{34}$ & $5.2 \times 10^{33}$ \\
Arches & $1.1 \times 10^4$ & 0.2 & $2 \pm 1$ & $1.1 \times 10^{38}$ 
& $1.5 \pm 0.2$ & 4 & $(2.0 - 4.0) \times 10^{34}$ & $9.3  \times 10^{34}$ \\
Quintuplet & $8.8 \times 10^3$ & 1 & $4 \pm 1$ & $4.4 \times 10^{38}$ 
& $2.4 \pm 0.5$ & 1 & $1.0 \times 10^{34}$ & $1.2 \times 10^{34}$ \\
\hline
\end{tabular}
\end{flushleft}
\end{table}  

The calculated X-ray luminosities are listed
in the last column of Table 1. They agree with the observed
values within a factor of 4 and do not show a systematic shift to the
smaller of larger values relative to the observed
luminosities. Clearly, the analysis of a larger sample of clusters
that includes more massive objects is required to avoid uncertainties
related with estimates of masses, ages and errors that result from the
deviation of star distributions from the idealized initial mass
function. For example, according to Leitherer \& Heckman (1995) 
the energy input rate from a coeval star cluster changes 
sharply in a short time interval between $2$~Myr and $4$~Myr. A 
small mistake in the age estimate then may change the predicted X-ray 
luminosity within a factor of 10 or more.  A smaller upper mass cutoff leads 
to the same uncertainty.
For a $\sim 10^4$\Msol \, cluster the relative dispersion of the 
observed X-ray luminosity due to the discreteness of the stellar population, 
$\sigma_X = [(L_X - \tilde{L}_X)^2]^{1/2} / \tilde{L}_X$, may reach $\sigma_X
\approx 3$ (see Figure 2 from Cervi\~no et al. 2002), and thus the
$L_X$ detected from a particular cluster may deviate from the mean
value within a factor of $\sim 3$. It seems therefore that more massive and 
more evolved star clusters are better candidates to compare with 
our model predictions.

Several more complications should be taken into consideration when comparing
the results from equation (\ref{eq.4}) to the observed X-ray emission.
In all young stellar clusters we have considered, the observed 
X-ray emission presents two components: a number of unresolved
point-like sources, probably associated with the most luminous individual
stars and local density enhancements resulting from the collisions
of nearby stellar winds (see Ozernoy et al. 1997, and the results
from Raga et al. 2001 and Rockefeller et al. 2004).
The contribution from the point-like sources varies from $\sim 20\%$
in the Quintuplet cluster to $\sim 80\%$ in NGC 3603 and cannot be
distinguished from the star cluster diffuse component in distant clusters.

After $\sim 3$ Myr the most massive stars leave the main sequence 
and explode as supernova. This results in the formation of X-ray binary
systems. The high mass X-ray binaries (HMXBs) contain a relativistic 
remnant of the supernova explosion (a neutron star or a black hole) which 
captures and accretes a wind from the secondary (O or B type), massive 
star. To transform a gravitational energy into the X-ray emission
effectively, it is essential that an accretion disc is formed or a
very strong magnetic field is present (see, for example, van Bever \& 
Vanbeveren, 2000, and references therein). Thus, it takes 4-5 Myr
for the HMXBs to become active, depending
on the upper mass limit occurring in the cluster. The HMXBs phase
is restricted by the life-time of the massive secondaries and is
typically few times $10^7$ yr. 
In the low mass X-ray binaries (LMXB) the optical component is a low
mass post-main-sequence star that fills its Roche lobe. 
In the instantaneous star formation LMXBs become active at later times
and certainly dominate the X-ray emission from older clusters
for which the SNe activity has terminated and the X-ray emission from zones 
A and B has vanished. 
LMXBs are often detected in globular clusters (see, for example, 
Maccarone et al. 2003, who found that in the NGC 4472, approximately 
40\% of the LMXBs are associated with the globular clusters and 
approximately 4\% of the globular clusters contain LMXBs). In the 
Milky Way around 10\% of the LMXBs are associated with globular 
clusters which contain less than 1\% of the stellar mass of the 
Galaxy (Liu et al., 2001). NGC 3603, Arches and Quituplet clusters 
are too young and do not contain active binaries.

The population synthesis models for massive star clusters which take
into consideration binary systems have been developed by a number of
groups (see, for example, Iben et al. 1995; Mas-Hesse \& Cervi\~no, 1999;
Dalton et al. 1995). Van Bever \& Vanbeveren (2000) have used such a
model to calculate the X-ray luminosity from the binary population of
instantaneous starbursts with different fractions of binaries
at the birth of the cluster. The HMXB component begins to contribute to
the X-ray emission after $\approx 4$ Myr. The normalized per one
solar mass X-ray emission does not depend on the initial mass of the 
cluster and soon reaches a value that varies between 
$10^{32}$ erg s$^{-1}$ M$^{-1}_{\odot}$ and $10^{33}$ erg s$^{-1}$  
M$^{-1}_{\odot}$ (see figures 1 and 2 from their paper).

In contrast to HMXBs, the normalized per unit solar mass X-ray 
luminosity from the thermalized star cluster plasma scales linearly 
with the star cluster mass: 
\begin{equation}
      \label{eq.10}
L_{X,A+B}/M_{SC} = 2.9 \times 10^{33} \Phi(T_{SC},Z_{SC}) M_6 /
                   R_{SC,1} T^3_{keV} ,
\end{equation}
where $M_6$ is the star cluster mass measured in units of $10^6$\Msol.
Figure 8 shows that for the most massive and compact young stellar 
clusters the X-ray luminosity of the thermalized star cluster plasma 
(zones A and B) may be comparable or even exceed that from the HMXBs.  
\begin{figure}[htbp]
\plotone{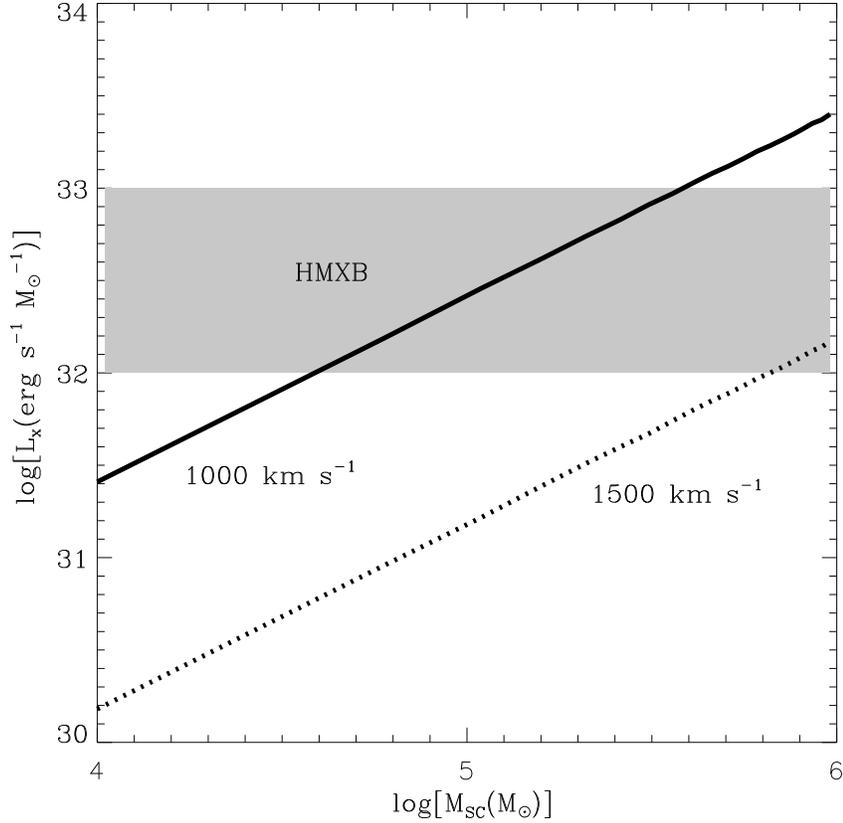}
\caption{X-ray luminosity of zones A and B per unit stellar mass 
as a function of the star cluster mass. The calculations have been
performed for a 1 pc clusters with 1000 km s$^{-1}$ ($T_{SC} = 0.95$ keV)
and 1500 km s$^{-1}$ ($T_{SC} = 2.1$ keV) terminal speeds indicated by 
the solid and dashed lines, respectively. The metallicity of the X-ray 
plasma is Solar. The shadow region on the diagram represents
the X-ray luminosity range expected from
the HMXB population (see van Bever \& Vanbeveren, 2000).}
\label{fig8}
\end{figure}

\section{Conclusions}

We have developed a simple way of estimating the X-ray emission 
generated by super star clusters. The derived expression 
(equation \ref{eq.4}) takes into consideration the intrinsic parameters
of SSCs and their winds ($R_{SC}, L_{SC}, V_{A,\infty}, Z_{SC}$). 
In particular, we have found that the expected X-ray luminosity 
from SSC is highly dependent on the star cluster wind terminal 
speed ($L_X \sim V^{-6}_{A,\infty}$), a quantity, related to the 
temperature of the hot thermalized plasma within the SSC volume,
and scales quadratically with the star cluster mass. 
The proposed relation seems to be in reasonable agreement with 
parameters of nearby clusters and their detected X-ray emissions.  

We have also compared the X-ray luminosity from the SSCs with the 
luminosity of the interstellar bubbles generated from the mechanical
interaction of the star cluster winds with the ISM. We 
found that the soft component and the total X-ray 
emission are usually dominated by the superbubble plasma.
The contribution from the SSC plasma may dominate  
only in the case of very massive ($\ge 10^6$\Msol) and compact 
(few parsecs) star clusters evolving into a low density ISM.
However the hard (2.0 keV - 8.0 keV) component of the X-ray emission
is usually dominated by the hot compact regions associated with the
SSCs. This implies that compact and massive star clusters should
be detected as hard compact X-ray sources embedded into extended 
regions of soft diffuse X-ray emission. 

On the other hand the comparison with the population synthesis models 
which take into consideration binary systems shows that the X-ray 
emission from the thermalized star cluster plasma normalized per 
unit stellar mass increases linearly with the mass of the cluster 
and may be comparable or even exceed that from the population of the 
HMXBs for the most compact and massive star clusters. Thus the 
thermalization of stellar winds and SNe ejecta, particularly in
massive, young and compact super star clusters, may present an 
X-ray production that is comparable or even larger than that expected
from the HMXB population.

\acknowledgments 
We thank E. Jim\'enez Bail\'on and M. Cervi\~no  for careful reading
of manuscript and helpful discussions. We also wish to express our 
thanks to our anonymous referee for a report full of suggestions to 
improve the paper.
This study has been supported by CONACYT - M\'exico, research grant 
47534-F and AYA2004-08260-CO3-O1 research grant from the Spanish 
Consejo Superior de Investigaciones Cient\'\i{}ficas. GTT acknowledges 
finantial support from the Secretar\'\i{}a de Estado de Universidades
e Investigaci\'on (Espa\~na) ref: SAB2004-0189 and the hospitality of 
the Instituto de Astrof\'\i{}sica de Andaluc\'\i{}a (IAA, CSIC) in 
Granada, Spain.





\end{document}